# A compositional ordering-driven morphotropic phase boundary in ferroelectric solid solutions

Yubai Shi,[1, 2] Yifan Shan,[1] Hongyu Wu,[1] Zhicheng Zhong,[3, 4] Ri He,[1, 2, *] Run-Wei Li[1, 2, 5, †]

[1]*CAS Key Laboratory of Magnetic Materials and Devices, Zhejiang Province Key Laboratory of Magnetic Materials and Application Technology, Ningbo Institute of Materials Technology and Engineering, Chinese Academy of Sciences, Ningbo 315201, China*

[2]*College of Materials Science and Opto-Electronic Technology, University of Chinese Academy of Sciences, Beijing 100049, China*

[3]*Department of Physics, University of Science and Technology of China, Hefei 230026, China*

[4]*Suzhou Institute for Advanced Research, University of Science and Technology of China, Suzhou 215123, China*

[5]*School of Future Technology, University of Chinese Academy of Sciences, Beijing 100049, China*

Ferroelectric solid solutions usually exhibit giant dielectric response and high piezoelectricity in the vicinity of the morphotropic phase boundary (MPB), where the structural phase transitions between the rhombohedral and the tetragonal phases as a result of the composition or strain variation. Here, we propose a compositional ordering-driven MPB in the specified compositional solid solutions. By preforming machine-learning potential based molecular dynamics simulations on lead zirconate titanate, we find a phase transition from the rhombohedral to tetragonal phase with the decrease of compositional ordering, leading to the MPB on temperature-ordering phase diagram. The compositional ordering-driven MPB can enhances the piezoelectricity with a magnitude comparable to that at the composition-driven MPB. Finally, we demonstrate that the mechanism of high piezoelectricity is polarization rotation driven by external field. This work provides an additional degree of freedom, compositional ordering, to design the high-performance piezoelectric materials.

---

[*]heri@nimte.ac.cn

[†]runweili@nimte.ac.cn

## I. INTRODUCTION

Morphotropic phase boundary (MPB) usually presents in phase diagram of ferroelectric and ferromagnetic solutions [1-8]. It has drawn constant interest due to the remarkable dielectric and magnetic properties exhibited. The essence of MPB is related to phase transition, generally displayed at the intersection area of tetragonal and rhombohedral phases in the phase diagram. There are two methods to generate the MPB of ferroelectric materials: changing composition ratio in solid solutions and mechanical pressure in pure compound. Lead zirconate titanate ($PbZr_{1-x}Ti_xO_3$, PZT) is a classical material for studying MPB [1]. In the composition phase diagram of PZT, the Ti-enriched and Zr-enriched areas are ferroelectric tetragonal and rhombohedral phases, respectively. The MPB is located at the transition region between the two phases, exhibiting high piezoelectricity. On the other hand, for a pure compound, such as lead titanate and bismuth ferrite, pressure could induce the phase transition from tetragonal to rhombohedral phase, forming MPB [6,9].

PZT is with typical $ABO_3$ perovskite structure, oxygen atoms forming the octahedra, of which Pb atoms occupy the interstices, and Zr/Ti atoms reside at the B-site, the center of the octahedra. The arrangement of B-site atoms can change, display a certain ordering. Variation of compositional ordering in solid solutions, or even in high entropy alloys, can change the electronic and atomic structures of the system, thereby causing effects on their mechanical, thermal, magnetic, and dielectric properties [10-12]. In the recent experiments, some advanced characterization techniques are available to determine the compositional ordering in solid solutions, such as atomic electron tomography and neutron/X-ray total scattering combined with reverse Monte Carlo method [13,14]. These results enable the possibility of studying compositional ordering in systems at the atomic scale. Based on these studies, compositional ordering variation in solid solutions could be considered to serve as a degree of freedom for inducing MPB. The arrangement of Zr/Ti atoms in PZT could potentially influence the piezoelectricity.

Using computational methods to study the compositional ordering can provide more insights for the subject. However, it is not feasible to study the order-disorder by density function theory (DFT), as large-scale systems are needed to simulate the disorder. The emergence of machine-learning-based potential functions makes it possible to overcome this limitation. Deep potential (DP) model, utilizing a deep neural network to capture the potential energy surface of material system from DFT calculations, could perform large-scale atomic dynamics simulations with DFT accuracy [15]. In recent work, we have developed a DP model for PZT, which can accurately reproduce the temperature-composition phase diagram and the piezoelectricity of PZT [16].

In this paper, we use the machine-learning-based DP model of PZT to explore the thermodynamic stability of PZT varying with the compositional ordering, and we demonstrate the decrease of ordering degree can cause a phase transition from tetragonal phase to rhombohedral phase. Furthermore, it is essential to highlight that the variation of compositional ordering can form the MPB in the temperature-ordering phase diagram, leading to high piezoelectricity. Finally, we reveal the mechanism of high piezoelectricity at the compositional ordering-driven MPB is polarization rotation. It is reasonable to assume that the phenomenon and mechanism can be widely applied to other ferroelectric solid solutions, such as hafnium zirconium oxide [17].

## II. COMPUTATIONAL METHODS

The machine-learning potential model is generated by DP-GEN [18]. It is currently the mainstream software for developing machine-learning potential models. In recent years, researchers study various material systems, especially perovskite ferroelectrics, based on the DP models and achieved good influence, proving the reliability of this strategy [16,19-24]. DP-GEN is a concurrent learning procedure control package using the deep neural network. DFT calculations needs to be used as the initial datasets. Then it enters the iterative process, training, labeling, and exploring the datasets. Finally, all datasets are collected and combined into the final dataset for a long-step training. The final generated DP model can effectively represent the PZT potential energy surface. For more details on the DFT parameters of the training set and accuracy test of DP model, please refer to the Methods section of our recent work [16].

The set of structures varying continuously from order to disorder is generated based on a strategy of random exchange. The basic ordered structure is a $20 \times 20 \times 20$ supercell consisting of 40,000 atoms. We randomly exchange the positions of two B-site atoms each step, and output a structure in 200 steps. After 20,000 steps, we continuously obtain 100 structures. We also use the method of random substitution to generate disordered structures for comparison. Using LAMMPS code, 100 structures of PbZr$_{0.5}$Ti$_{0.5}$O$_3$ are generated by different random seeds [25,26]. Here, the ordering degree σ of the structure and the configurational entropy $S_{conf}$ contributed by Zr/Ti atoms are defined based on Bragg-William approximation. We define the sites of Zr atoms in the ordered arrangement as $α$ sublattice. And the σ and $S_{conf}$ are specifically calculated by the following equations:

$$\sigma = \left|(N_\alpha^{Zr} - N_\alpha^{Ti})/\frac{N}{2}\right|.$$

$$S_{conf} = -\frac{k_B N}{2}[(1+\sigma)ln(1+\sigma) + (1-\sigma)ln(1-\sigma) + 2ln2].$$

Where $N_\alpha^{Zr}$ is the number of Zr atoms occupied the $α$ sublattice, $N_\alpha^{Ti}$ the number of Ti atoms occupied the $α$ sublattice, and $N$ is the total number of Zr/Ti atoms. Because the key point we focus on is the order-disorder variation of PZT, the configurational entropy plays an unneglectable role in analyzing the thermodynamic stability of structures. In this work, only total energy and configurational entropy are considered in free energy:

$$F = U - TS_{conf}.$$

$U$ can be obtained by the kinetic and potential energies output by LAMMPS.

Based on the DP model and prepared initial structures, we conducted molecular dynamics simulations. calculations are performed by LAMMPS code with periodic boundary conditions and at standard pressure [25,26]. We set the temperature range from 100K to 300K. The time step is set to 0.001 ps. At a specified temperature, the equilibrium run is 40 ps. The simulation environment is set as a triclinic box, allowing for full relaxation of lattice constants and lattice angles. Applying an electric field to the system is achieved by applying an external force to the atoms as an equivalent method. The forces acting on different atoms are determined by multiplying the electric field by the reference Born effective charge [27]: $Z_{Pb}^* = 3.90$, $Z_{Zr}^* = 5.85$, $Z_{Ti}^* = 7.06$. To keep the electric neutrality and take into account the solid solution characteristics of PZT, a dynamic average adjustment scheme is applied to the Born effective charge of oxygen atoms: $Z_O^* = -[Z_{Pb}^* + xZ_{Ti}^* + (1-x)Z_{Zr}^*]/3$, where $x$ represents the concentration of Ti. The piezoelectric coefficient $d_{33}$ is obtained based on: $d_{33} = (\partial x_3/\partial E_3)_X$, where $E_3$ is external electric field, $x_3$ the strain induced, 3 the along the $z$ axis, and $X$ represents the external stress.

### III. RESULTS AND DISCUSSIONS

In our recent work, the disordered arrangements of B-site atoms in the PZT are considered [16]. However, the compositional ordering could have a significant impact on the thermodynamic stability of structure, which further influences piezoelectricity. Firstly, for the composition with PbZr$_{0.5}$Ti$_{0.5}$O$_3$, we construct three typical ordered structures as shown in Fig. 1, as well as the different disordered structures. The energy fluctuation caused by different disordered arrangements is within 0.03 meV/atom, demonstrating the negligible impact of different disordered structures on their energy. We summarize results of various arrangements in Table I. It is shown that the ordered types are more energy favorable than the disordered types, and the C-type has the lowest energy. PZT in different ordered types exhibit different ground state structures. We further statistically analyze the lattice information (see Table II). Calculated results show that the ground state of PbZr$_{0.5}$Ti$_{0.5}$O$_3$ in disordered type is the rhombohedral phase which our recent work also pointed out, while A-type exhibits the monoclinic phase, C-type the tetragonal phase, and G-type the rhombohedral phase. For these ground state structures, we also calculate the phonon spectrum (see Fig. 2). It is shown that there are no lattice vibration modes with frequencies less than 0, demonstrating the stabilities of these ground state structures.

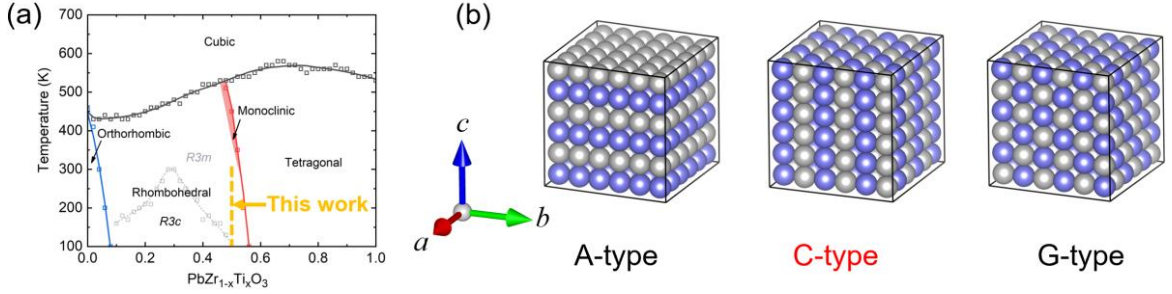

FIG. 1. (a) DP-predicted temperature-composition phase diagram of PZT, and the red line is MPB. [16]. The yellow dash line represents the composition of PZT discussed in this paper, which is 0.5, to the left of the MPB. (b) Three types of ordered arrangement of B-site atoms. The highlighted C-type serves as the basis for the subsequent discussions. The blue spheres represent Zr atoms, while the gray spheres represent Ti atoms. The schematic diagram excludes Pb and O atoms.

TABLE I. The $\Delta E$ between $PbZr_{0.5}Ti_{0.5}O_3$ in different arrangement types and C-type after atomic relaxation, calculated by DP and DFT, respectively.

| Arrangement Type | $\Delta E^{DP}$ (meV/atom) | $\Delta E^{DFT}$ (meV/atom) |
|---|---|---|
| A-type | 5.67 | 5.11 |
| **C-type** | **0** | **0** |
| G-type | 4.41 | 4.80 |
| Disordered | 6.61±0.03 | |

TABLE II. Structure information of the pseudocubic unit cell of $PbZr_{0.5}Ti_{0.5}O_3$ in different arrangement types after atomic relaxation, calculated by DP and DFT, respectively.

| Arrangement Type | Lattice Constants (DP) (Å) | Lattice Angles (DP) (Å) | Lattice Constants (DFT) (Å) | Lattice Angles (DFT) (Å) |
|---|---|---|---|---|
| A-type | $a = 4.091$<br>$b = 4.091$<br>$c = 4.002$ | $\alpha = 89.66°$<br>$\beta = 89.66°$<br>$\gamma = 89.23°$ | $a = 4.087$<br>$b = 4.087$<br>$c = 3.999$ | $\alpha = 89.76°$<br>$\beta = 89.76°$<br>$\gamma = 89.26°$ |
| C-type | $a = 3.989$<br>$b = 3.989$<br>$c = 4.254$ | $\alpha = 90.00°$<br>$\beta = 90.00°$<br>$\gamma = 90.00°$ | $a = 3.983$<br>$b = 3.983$<br>$c = 4.261$ | $\alpha = 90.00°$<br>$\beta = 90.00°$<br>$\gamma = 90.00°$ |
| G-type | $a = 4.046$<br>$b = 4.046$<br>$c = 4.046$ | $\alpha = 89.57°$<br>$\beta = 89.57°$<br>$\gamma = 89.57°$ | $a = 4.041$<br>$b = 4.041$<br>$c = 4.041$ | $\alpha = 89.75°$<br>$\beta = 89.75°$<br>$\gamma = 89.75°$ |
| Disordered | $a = 4.059±0.003$<br>$b = 4.058±0.002$<br>$c = 4.058±0.003$ | $\alpha = 89.49±0.01$<br>$\beta = 89.49±0.01$<br>$\gamma = 89.49±0.01$ | | |

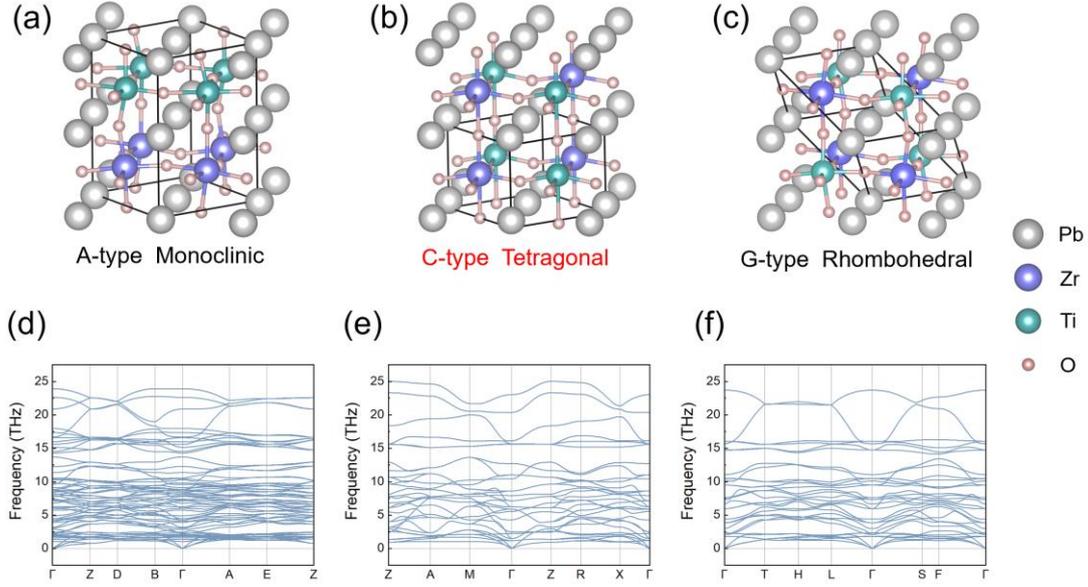

FIG. 2. Primitive cell of PbZr$_{0.5}$Ti$_{0.5}$O$_3$ of ground state in (a) A-type, (b) C-type, and (c) G-type, exhibiting monoclinic, tetragonal, and rhombohedral phase, respectively. Corresponding phonon spectrum of (c) A-type, (d) C-type, and (e) G-type.

Considering the ordered C-type structure is the tetragonal phase, while the disordered structure is the rhombohedral phase, what remains to be clarified is whether the compositional ordering could serve a degree of freedom, like composition ratio and pressure, leading a rhombohedral to tetragonal phase transition and inducing the MPB. To explore this issue, we design a set of structures continuously varying from ordered C-type to disordered type by exchange the sites of Zr and Ti atoms randomly. We quantify the ordering degree ($\sigma$) of the structures, where $\sigma = 1$ represents prefect ordered C-type one, and $\sigma = 0$ the disordered type one. As shown in Fig. 3(a), with an increase in the exchange count, $\sigma$ decreases. $\sigma$ converge to 0 after 9000 exchanges approximately. While for randomly generated disordered structures, $\sigma$ consistently fluctuates around 0 with negligible variations. The results demonstrate that our random exchange strategy effectively provides a continuous structural variation from the C-type to disordered structure.

We also compute the total energy ($U$) of these structures at 100 K. Herein, we used the energy differences between different structures and the C-type structure to denote the thermodynamic stability. In Fig. 3(b), it shows that as $\sigma$ decreases, the total energy gradually increases. As the ordering degree of the structure approximately less than 0.33, the total energy converges to the same value with the randomly generated disordered structures. It is necessary to consider configurational entropy at finite temperatures. We calculated the free energy ($F$) by adding the contribution of configurational entropy to the total energy (see method). In Fig. 3(b), The results show that the free energy also displays similar behavior to the total energy. However, the introduction of configurational entropy leads to a greater impact for more disordered structures. It reduces the energy difference between disordered and ordered structures. Therefore, as σ decreases, the free energy converges more quickly to the same value as the disordered structures, compared to the total energy. The above results indicate that configurational entropy does not make a qualitative impact on the results.

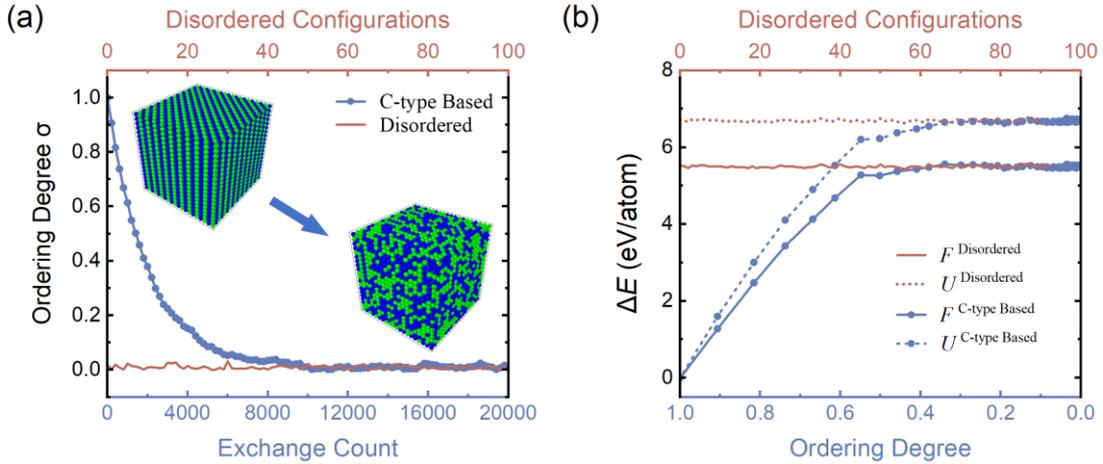

FIG. 3. (a) The ordering degree σ of different structures. The blue line represents the set of structures varying from the ordered C-type to the disordered type by random exchange, while the red line represents the disordered structures generated by different random seeds. (b) Total energy ($U$) and free energy ($F$) differences between different structures and the ordered C-type, at 100 K. Solid lines represent the free energies and the dash lines represent the total energies. The set of structures varying from the ordered C-type to disordered are denoted by σ.

In addition to affecting the energy, compositional ordering also causes changes in the structure. we plotted the lattice variation of a 40000-atom supercell with decreasing the σ (see Fig. 4). When σ > 0.5, PZT is the tetragonal phase, where the *a* and *b*-axis are equal and less than the *c*-axis, and the three axes are orthogonal. When the σ reduces to below 0.5, the lattice transitions from the tetragonal phase to the rhombohedral phase. When the σ further decreases to 0.2, PZT is the rhombohedral phase. We elevate the temperature and get similar behaviors. In Fig. 4(c), We plot the temperature-ordering phase diagram of PbZr$_{0.5}$Ti$_{0.5}$O$_3$ using lattice angles as variables. On the side of tiny σ, indicating the disordered arrangement of B-site atoms, it corresponds to the rhombohedral phase. This is consistent with our recent findings [16]. It is noteworthy that the phase transition from tetragonal to rhombohedral phase induced by compositional ordering is continuous, different from the abrupt changed one induced by composition variation. Therefore, as the temperature increases, the boundary between the rhombohedral and tetragonal phases slightly shifts to the right, forming a diffuse MPB. This indicates that the compositional ordering can serve as a degree of freedom to induce the MPB. Additionally, different from alloys, in the thermodynamic equilibrium process, there is no atomic diffusion in the lattice of PZT, with the increase of temperature. This ensures the uniqueness of ordering degree in each structure.

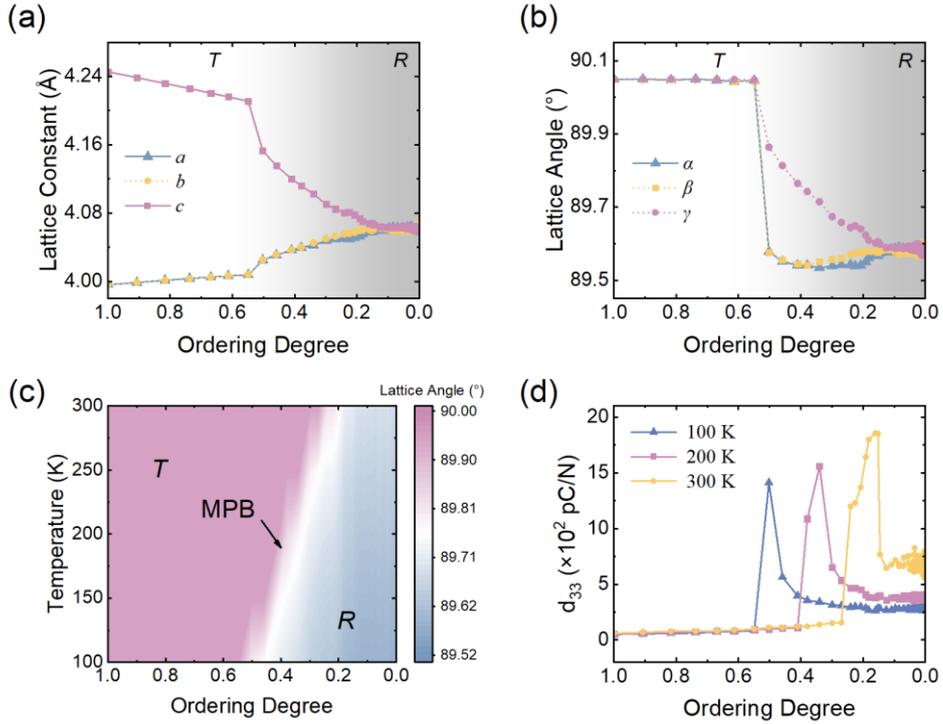

FIG. 4. (a) Lattice constants and (b) lattice angles of PbZr$_{0.5}$Ti$_{0.5}$O$_3$ as functions of σ at 100 K. The gradient from white to gray represents the transition from the tetragonal (T) phase to the rhombohedral (R) phase. (c) Temperature-ordering phase diagram of PbZr$_{0.5}$Ti$_{0.5}$O$_3$ at intervals of 50 K, ranging from 100 K to 300 K. σ = 1 represents the structure in ordered C-type. The white transition region represents the boundary between the tetragonal phase and rhombohedral phase, known as the MPB. (d) The variation of $d_{33}$ with respect to σ at different temperatures. $d_{33}$ reaches its maximum at 0.5, 0.33, and 0.16, respectively.

To explore the nature of the compositional ordering-driven MPB, we calculate piezoelectric coefficient $d_{33}$. Based the derivative of strain with respect to electric field with range of 0~1 ×10$^2$ kV/cm (for details, see methods), we calculate $d_{33}$ as shown in Fig. 4(d). $d_{33}$ exhibits a sharp peak at different temperatures, the peak values are all around thousands of pC/N, comparable to those at the composition-driven MPB [16]. As the temperature increases, the peak of $d_{33}$ gradually shift to right. This corresponds to the decrease in σ at the MPB in the phase diagram as the temperature rises. Therefore, MPB is evidently the region with the best piezoelectric performance. The behavior of piezoelectricity at the compositional ordering-driven MPB is similar to those at the composition and pressure-driven MPB in previous studies [1,6].

To unravel the mechanism of the high piezoelectricity caused by the compositional ordering-driven MPB, we analyze the frequency distribution on the displacement of B-site atoms relative to the oxygen octahedral center. We presented the results for three structures, corresponding to the ordered structure (σ = 1), the partially ordered structure at the MPB (σ = 0.5), and the disordered structure (σ = 0). As shown in Fig. 5, When no electric field is applied, the displacement components of B-site atoms in the ordered structure are 0 along the *x/y*-axis, while the *z*-axis component exhibits two peaks, representing Zr and Ti, respectively. The splitting of the displacement is due to the larger Born effective charge and smaller relative atomic mass of Ti, in contrast with Zr. The polarization of the ordered structure is clearly oriented along the [001] direction under zero electric field, aligning with the tetragonal symmetry. While for σ = 0, the displacements along the three axes are equal, and the two peaks of Zr and Ti merge, clearly indicating the polarization orientation as [111], aligning with the rhombohedral symmetry. For σ = 0.5, it exhibits a transitional state between the tetragonal and rhombohedral phases. Specifically, there is a non-zero displacement along the *x/y*-axis, but it is not equal to the one along the *z*-axis. As a relatively small electric field of 1×10$^2$ kV/cm is applied, the displacements at σ = 0 and 1 show almost unchanged as shown inset of Fig. 5, indicating such small electric field is not sufficient to change the polarization orientation. However, the structure with σ = 0.5 shows significant changes with the same electric field, with the displacement along the *x/y*-axis reducing to zero and the *z*-axis continuing to increase, resulting in the polarization rotation to [001] direction. This clarify the reason why the piezoelectric performance at the MPB is better than the other region. Therefore, the high piezoelectricity induced by electric field-driven polarization rotation at the compositional ordering-driven MPB is the same as those at the composition and pressure-driven MPB [6,16].

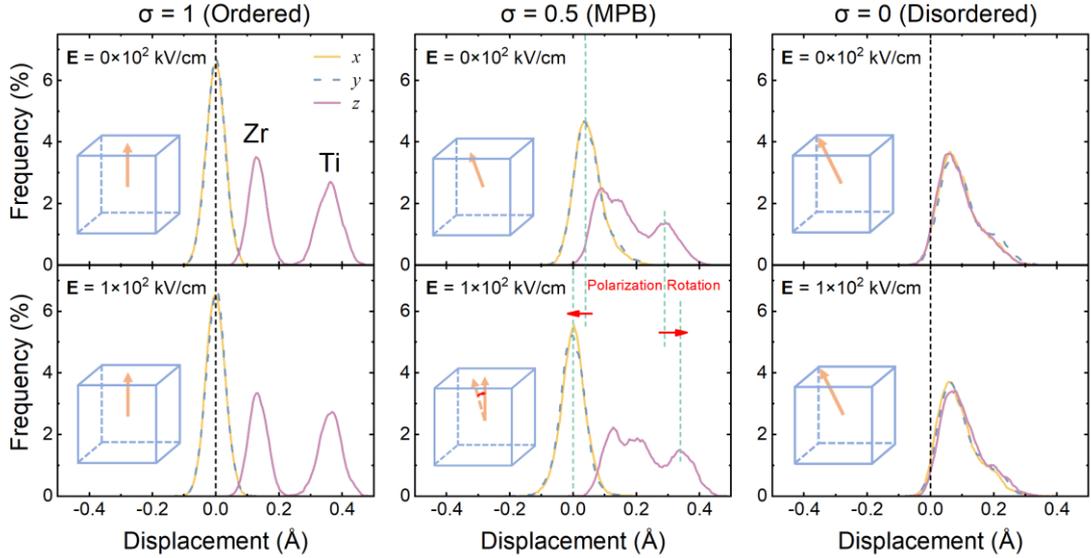

FIG. 5. Frequency distributions of B-site atoms relative to the oxygen octahedral center displacements at 100 K. We analyze the three structures under **E** = 0 and $1\times10^2$ kV/cm, in σ = 1, 0.5, and 0, respectively.

## IV.    CONCLUSION

In summary, we consider compositional disordered and three ordered types of PbZr$_{0.5}$Ti$_{0.5}$O$_3$. It is found that PZT in ordered C-type is the most energy favorable. Based on this result, we design a set of structures transitioning continuously from the ordered C-type to the disordered. Then, we investigate the thermodynamic and piezoelectric properties of PZT experiencing the order-disorder variation. In the temperature-ordering phase diagram, as ordering degree (σ) decreases, the structure transitions from the tetragonal phase to the rhombohedral phase, forming the MPB. The curve of the piezoelectric coefficient $d_{33}$ with respect to σ exhibits a sharp peak at the MPB, shifting to the right with increasing temperature. Finally, we analyze the displacement distribution of B-site atoms to demonstrate that the high piezoelectric coefficient is induced by polarization rotation, consistent with composition and pressure-driven MPB. We believe that the compositional ordering-driven MPB is a universal phenomenon in ferroelectric solid solutions. This work can be helpful to design and synthesize high-performance piezoelectric and dielectric energy storage materials by controlling the compositional ordering of the complex solid solutions in experiments.

## ACKNOWLEDGMENTS

This work was supported by the National Key R&D Program of China (Grants No. 2021YFA0718900, No. 2022YFA1403000, and No. 2021YFE0194200), the Key Research Program of Frontier Sciences of CAS (Grant No. ZDBS-LY-SLH008), the National Natural Science Foundation of China (Grants No. 11974365, No. 12204496, and No. 12161141015), the K.C. Wong Education Foundation (Grant No. GJTD-2020–11), the Science Center of the National Science Foundation of China (Grant No. 52088101), and the Zhejiang Provincial Natural Science Foundation (No. Q23A040003).

**Reference**

[1] B. Jaffe, W. R. Cook, and H. L. Jaffe, *Piezoelectric Ceramics* (Academic Press, 1971).
[2] B. Noheda, D. Cox, G. Shirane, J. Gonzalo, L. Cross, and S. Park, A monoclinic ferroelectric phase in the $Pb(Zr_{1-x}Ti_x)O_3$ solid solution, Applied Physics Letters **74**, 2059 (1999).
[3] R. Guo, L. Cross, S. Park, B. Noheda, D. Cox, and G. Shirane, Origin of the high piezoelectric response in $PbZr_{1-x}Ti_xO_3$, Physical Review Letters **84**, 5423 (2000).
[4] H. Fu and R. E. Cohen, Polarization rotation mechanism for ultrahigh electromechanical response in single-crystal piezoelectrics, Nature **403**, 281 (2000).
[5] T. Asada and Y. Koyama, Ferroelectric domain structures around the morphotropic phase boundary of the piezoelectric material $PbZr_{1-x}Ti_xO_3$, Physical Review B **75**, 214111 (2007).
[6] M. Ahart, M. Somayazulu, R. Cohen, P. Ganesh, P. Dera, H.-k. Mao, R. J. Hemley, Y. Ren, P. Liermann, and Z. Wu, Origin of morphotropic phase boundaries in ferroelectrics, Nature **451**, 545 (2008).
[7] S. Yang, H. Bao, C. Zhou, Y. Wang, X. Ren, Y. Matsushita, Y. Katsuya, M. Tanaka, K. Kobayashi, and X. Song, Large magnetostriction from morphotropic phase boundary

in ferromagnets, Physical Review Letters **104**, 197201 (2010).

[8] X. Ke, C. Zhou, B. Tian, Y. Matsushita, X. Ren, S. Yang, and Y. Wang, Direct evidence of magnetization rotation at the ferromagnetic morphotropic phase boundary in $Tb_{1-x}Dy_xFe_2$ system, Physical Review B **108**, 224419 (2023).

[9] R. Zeches, M. Rossell, J. Zhang, A. Hatt, Q. He, C.-H. Yang, A. Kumar, C. Wang, A. Melville, and C. Adamo, A strain-driven morphotropic phase boundary in $BiFeO_3$, Science **326**, 977 (2009).

[10] F. C. Nix and W. Shockley, Order-disorder transformations in alloys, Reviews of Modern Physics **10**, 1 (1938).

[11] Y. Zhang, L. Chen, H. Liu, S. Deng, H. Qi, and J. Chen, High‐performance ferroelectric based materials via high-entropy strategy: Design, properties, and mechanism, InfoMat **5**, e12488 (2023).

[12] H. Liu, X. Shi, Y. Yao, H. Luo, Q. Li, H. Huang, H. Qi, Y. Zhang, Y. Ren, and S. D. Kelly, Emergence of high piezoelectricity from competing local polar order-disorder in relaxor ferroelectrics, Nature Communications **14**, 1007 (2023).

[13] S. Moniri, Y. Yang, J. Ding, Y. Yuan, J. Zhou, L. Yang, F. Zhu, Y. Liao, Y. Yao, and L. Hu, Three-dimensional atomic structure and local chemical order of medium- and high-entropy nanoalloys, Nature **624**, 564 (2023).

[14] H. Liu, Z. Sun, J. Zhang, H. Luo, Y. Yao, X. Wang, H. Qi, S. Deng, J. Liu, and L. C. Gallington, Local Chemical Clustering Enabled Ultrahigh Capacitive Energy Storage in Pb-Free Relaxors, Journal of the American Chemical Society **145**, 19396 (2023).

[15] L. Zhang, J. Han, H. Wang, R. Car, and E. Weinan, Deep potential molecular dynamics: a scalable model with the accuracy of quantum mechanics, Physical Review Letters **120**, 143001 (2018).

[16] Y. Shi, R. He, B. Zhang, and Z. Zhong, Revisiting the phase diagram and piezoelectricity of lead zirconate titanate from first principles, Physical Review B **109**, 174104 (2024).

[17] M. H. Park, Y. H. Lee, H. J. Kim, Y. J. Kim, T. Moon, K. D. Kim, S. D. Hyun, and C. S. Hwang, Morphotropic phase boundary of $Hf_{1-x}Zr_xO_2$ thin films for dynamic random access memories, ACS Applied Materials & Interfaces **10**, 42666 (2018).

[18] Y. Zhang, H. Wang, W. Chen, J. Zeng, L. Zhang, H. Wang, and E. Weinan, DP-GEN: A concurrent learning platform for the generation of reliable deep learning based potential energy models, Computer Physics Communications **253**, 107206 (2020).

[19] R. He, H. Wu, L. Zhang, X. Wang, F. Fu, S. Liu, and Z. Zhong, Structural phase transitions in $SrTiO_3$ from deep potential molecular dynamics, Physical Review B **105**, 064104 (2022).

[20] H. Wu, R. He, Y. Lu, and Z. Zhong, Large-scale atomistic simulation of quantum effects in $SrTiO_3$ from first principles, Physical Review B **106**, 224102 (2022).

[21] R. He, B. Zhang, H. Wang, L. Li, T. Ping, G. Bauer, and Z. Zhong, Ultrafast switching dynamics of the ferroelectric order in stacking-engineered ferroelectrics, Acta Materialia **262**, 119416 (2024).

[22] R. He, H. Xu, P. Yang, K. Chang, H. Wang, and Z. Zhong, Ferroelastic Twin-Wall-Mediated Ferroelectriclike Behavior and Bulk Photovoltaic Effect in $SrTiO_3$, Physical

Review Letters **132**, 176801 (2024).

[23] J. Wu, J. Yang, L. Ma, L. Zhang, and S. Liu, Modular development of deep potential for complex solid solutions, Physical Review B **107**, 144102 (2023).

[24] J. Wu, J. Yang, Y.-J. Liu, D. Zhang, Y. Yang, Y. Zhang, L. Zhang, and S. Liu, Universal interatomic potential for perovskite oxides, Physical Review B **108**, L180104 (2023).

[25] S. Plimpton, Fast parallel algorithms for short-range molecular dynamics, Journal of Computational Physics **117**, 1 (1995).

[26] A. P. Thompson, H. M. Aktulga, R. Berger, D. S. Bolintineanu, W. M. Brown, P. S. Crozier, P. J. in't Veld, A. Kohlmeyer, S. G. Moore, and T. D. Nguyen, LAMMPS- a flexible simulation tool for particle-based materials modeling at the atomic, meso, and continuum scales, Computer Physics Communications **271**, 108171 (2022).

[27] W. Zhong, R. King-Smith, and D. Vanderbilt, Giant LO-TO splittings in perovskite ferroelectrics, Physical Review Letters **72**, 3618 (1994).